\documentclass[preprintnumbers, floatfix, preprintnumbers, letterpaper,onecolumn, superscriptaddress,nofootinbib,11pt]{revtex4-2}
\pdfoutput=1
\usepackage{graphicx}
\usepackage{microtype}
\usepackage{amsmath}
\usepackage{amssymb}
\usepackage{subfigure}
\usepackage{url}
\usepackage{hyperref}
\usepackage{bigints}
\usepackage{xcolor}
\usepackage{color}
\usepackage{mathrsfs}
\usepackage{calrsfs}
\usepackage{amsfonts}
\usepackage{tabularx}
\usepackage{latexsym}
\usepackage{ragged2e}
\usepackage{textcomp}
\usepackage{float}

\usepackage{caption}
\definecolor{amaranth}{rgb}{0.9, 0.17, 0.31}
\DeclareCaptionJustification{justified}{\leftskip=0pt \rightskip=0pt \parfillskip=0pt plus 1fil}
\hypersetup{linktoc=all,
	colorlinks, linkcolor={blue},
	citecolor={blue}, urlcolor={amaranth}}

\begin{document}
\title{The stochastic gravitational wave background from QCD phase transition in the framework of higher-order GUP}

\author{Zhong-Wen Feng}
\altaffiliation{Email: zwfengphy@cwnu.edu.cn}
\affiliation{School of Physics and Astronomy, China West Normal University, Nanchong 637009, China}
\author{Long-Xiang Li}
\affiliation{School of Physics and Astronomy, China West Normal University, Nanchong 637009, China}
\author{Shi-Yu Li}
\affiliation{School of Physics and Astronomy, China West Normal University, Nanchong 637009, China}
\author{Qing-Quan Jiang}
\affiliation{School of Physics and Astronomy, China West Normal University, Nanchong 637009, China}
\author{Xia Zhou}
\affiliation{School of Physics and Astronomy, China West Normal University, Nanchong 637009, China}

\begin{abstract}
This work studies the impact of a new higher-order generalized uncertainty principle (GUP) on the stochastic gravitational wave background (SGWB) associated with a QCD-scale first-order phase transition. Assuming a strongly first-order transition at the QCD-scale as a phenomenological benchmark, the analysis shows that the sign and magnitude of the dimensionless deformation parameter $\beta_0$ play a crucial role. For negative $\beta_0$, the thermodynamic quantities of the radiation fluid develop a maximal temperature beyond which entropy and pressure vanish, and the SGWB spectrum exhibits divergent behavior at high temperatures, so this branch is discarded as phenomenologically inconsistent. For positive $\beta_0$, the higher-order GUP shifts the SGWB peak frequency towards lower values and slightly enhances the peak energy density, with the size of the effect controlled by $\beta_0$. For natural values $\beta_0=\mathcal{O}\left( 1 \right)$ the corrections at QCD temperatures are strongly suppressed, whereas larger benchmark values still compatible with existing experimental and cosmological bounds can induce appreciable shifts in the SGWB spectrum. A future detection of a QCD-scale first-order SGWB would therefore allow the framework developed here to be used to translate the measured signal into constraints on the higher-order GUP parameter, providing an indirect probe of quantum gravity effects.
\end{abstract}

\maketitle
\section{Introduction}
\label{intro}
The first direct detection of gravitational waves (GWs) by the LIGO collaboration in 2015 marked a milestone in modern physics, it confirmed
a key prediction of general relativity and opened the era of gravitational-wave astronomy, allowing violent cosmic events to be observed directly through spacetime fluctuations. Subsequent detections by the LIGO and Virgo collaborations have revealed a variety of GWs sources, including binary black hole mergers~\cite{LIGOScientific:2016aoc}, binary neutron star coalescences~\cite{LIGOScientific:2017vwq} and black hole-neutron star systems~\cite{LIGOScientific:2021qlt}. These discoveries have greatly stimulated theoretical and observational interest in GWs and their implications for cosmology and fundamental physics.

In addition to individually resolved compact-binary signals, there has been growing attention to the stochastic gravitational wave background (SGWB), which consists of the incoherent superposition of GWs generated by a large number of unresolved astrophysical and cosmological sources~\cite{Maggiore:1999vm,Christensen:2018iqi,Bian:2024bzl}. On the cosmological side, phase transitions in the early Universe provide a particularly intriguing class of potential SGWB sources. In particular, strongly first-order phase transitions at the electroweak temperature scale ($T\sim 100~{\rm GeV}$) and at temperatures of order the Quantum Chromodynamics (QCD) scale ($T\sim 0.1~{\rm GeV}$) can produce GWs through bubble collisions, long-lived sound waves in the plasma and magnetohydrodynamic turbulence~\cite{Kosowsky:1991ua,Kosowsky:1992rz,Huber:2008hg,Hindmarsh:2013xza,Hindmarsh:2015qta,Kamionkowski:1993fg,Binetruy:2012ze}. The resulting background is expected to populate the nanohertz (nHz) frequency band, which coincides with the sensitivity windows of pulsar timing array (PTA) experiments such as SKA~\cite{Hall:2009}, IPTA~\cite{Manchester:2013ndt}, EPTA~\cite{Kramer:2013kea}, NANOGrav~\cite{NANOGrav:2023hde} and CPTA~\cite{Xu:2023wog}. Recent PTA data already show tantalizing evidence for a common-spectrum process consistent with a SGWB of astrophysical or cosmological origin~\cite{Zhao:2024yau}, strengthening the motivation to investigate in detail how various high-energy phenomena might be imprinted in the SGWB.

Since a cosmological SGWB is typically generated in the very early universe, when temperatures, energy densities, and curvature scales are extremely high, it offers one of the observational windows in which quantum gravity (QG) effects might leave an observable imprint.  One of the most studied phenomenological approaches to QG is the generalized uncertainty principle (GUP), which modifies the Heisenberg uncertainty relation by introducing a minimal length scale, typically of the order of the Planck length~\cite{Gialamas:2024ivq}. This minimal length is suggested by several candidate theories of QG, including string theory and loop quantum gravity~\cite{Garay:1994en,Amelino-Camelia:2000stu}. The GUP implies deformed commutation relations between position and momentum, which in turn can modify the momentum distribution and thermodynamic properties of matter at very high energies or densities. In an early Universe context, the GUP corrections can alter the thermodynamics of a relativistic plasma, for instance by deforming the black-body spectrum and changing the entropy and equation of state of a radiation fluid~\cite{Tawfik:2015rva}. These changes can affect the expansion history and the redshift of cosmological SGWB sources, thereby providing an indirect probe of QG effects. Work in this area has already been initiated. In Refs.~\cite{Moussa:2021qlz,Moussa:2021gxb} Moussa \textit{et al.} incorporated two widely studied GUP models into the thermodynamics of the QCD plasma and showed that the resulting modifications can shift the peak frequency and amplitude of a SGWB generated by a QCD-scale first-order phase transition. One of the GUP models considered there is the Kempf–Mangano–Mann (KMM) form $\Delta x \Delta p \ge \frac{\hbar}{2}[1 + \beta_0 \ell_p^2 (\Delta p)^2/\hbar^2]$, while the other is the Ali–Das–Vagenas (ADV) model with linear and quadratic terms in the momentum uncertainty. These ``low-order" GUP models capture some QG-inspired features, but they also suffer from some limitations. First, the traditional forms typically involve only the leading quadratic (and possibly linear) terms in momentum, and therefore their applicability is restricted to momenta well below the Planck scale. At higher energies, higher-order terms are expected to become important and may change the phenomenology. Second, the standard GUP formulations do not naturally incorporate a maximal momentum, in contrast to frameworks such as doubly special relativity (DSR) in which both a minimal length and a maximal momentum arise. Third, in many setups the sign of the GUP parameter is fixed to be positive, yet in other contexts, most notably in the study of compact stars, the negative GUP parameters have been argued to resolve apparent conflicts with observations. For example, a positive deformation parameter can destroy the Chandrasekhar limit for white dwarfs and neutron stars, while a negative parameter restores a physically acceptable limiting mass~\cite{Rashidi:2015rro,Ong:2018zqn}. The negative GUP parameters have also been linked to non-trivial spacetime structures and emergent crystalline-like universes~\cite{Buoninfante:2019fwr,Jizba:2009qf}. These issues have triggered renewed interest in higher-order and sign-reversed GUP models and their implications~\cite{Pedram:2011gw,Shababi:2017zrt,Chung:2019raj,Hassanabadi:2019eol,Petruzziello:2020een,Zhao:2020xmp,Bosso:2023aht}.

To address these shortcomings, Pedram proposed a non-perturbative, higher-order GUP that remains well defined at arbitrarily high momenta and incorporates a maximal momentum in a way consistent with DSR~\cite{Pedram:2012my}. More recently, Du and Long introduced a new higher-order GUP (Du–Long) model in which the deformation depends on position uncertainty rather than momentum uncertainty~\cite{Du:2022mvr}. In this framework, the uncertainty relation takes the form as follows
\begin{align}
\label{eq1}
\Delta x\Delta p \ge \frac{\hbar }{2}\frac{1}{{1 + \left( {{{16\beta } \mathord{\left/ {\vphantom {{16\beta } {\Delta {x^2}}}} \right.
 \kern-\nulldelimiterspace} {\Delta {x^2}}}} \right)}},
\end{align}
where $\Delta x$ and $\Delta p$ denote the position and momentum uncertainties, respectively, and the deformation parameter is written as $\beta = {{\beta _0}\ell _p^2}={{{\beta _0}} \mathord{\left/  {\vphantom {{{\beta _0}} {m_p^2}}} \right.  \kern-\nulldelimiterspace} {m_p^2}}$ with  a dimensionless GUP parameter $\beta_0$, the Planck length  $\ell_p$ and the Planck mass $m_p$. This model possesses several interesting features. It admits both positive and negative values of $\beta_0$ while yielding a unified minimal length $\Delta x_{\min} = 4\sqrt{|\beta|} = 4\sqrt{|\beta_0|}\,\ell_p$ that is independent of the sign of the parameter. For $\beta_0>0$ a maximal momentum $\Delta p_{\max} = \hbar / (16\sqrt{|\beta|})$ arises, in line with the expectations of higher-order GUP models. Moreover, the minimal length is model independent and can be interpreted as a universal QG cutoff. These properties make the Du–Long GUP an attractive candidate for investigating the impact of higher-order and sign-reversed GUP corrections in cosmology. On this basis, several recent studies have explored its implications for cosmological dynamics, phase transitions, and nucleosynthesis, finding that the model can significantly modify the evolution of the early Universe and the associated thermodynamic quantities~\cite{Roushan:2024fog,Feng:2024zor,Luo:2023rhk,Luo:2024vdd}.

An important conceptual point concerns the nature of the QCD transition itself. At vanishing baryon chemical potential, state-of-the-art lattice QCD simulations indicate that the thermal QCD transition is a crossover rather than a strongly first-order phase transition. In this case, the Standard Model does not predict a SGWB sourced by a genuine first-order QCD transition. However, a strongly first-order transition at QCD temperatures can arise in various extensions of the Standard Model or at finite baryon density (see, e.g., \cite{Aoki:2006we,Bhattacharya:2014ara,Schwaller:2015tja,Balazs:2016tbi,Dev:2016feu}), and the corresponding GW spectra have been widely used as phenomenological templates in the literature~\cite{Caprini:2015zlo,Amaro-Seoane:2012vvq,Ruan:2018tsw}. Moreover, even independently of a specific microphysical realization, a QCD-scale first-order transition provides a convenient theoretical laboratory for studying how new physics could affect a nanohertz SGWB. The present work adopts this phenomenological perspective: a strongly first-order transition at the QCD scale is assumed as a working hypothesis, and the focus is placed on quantifying how the Du–Long GUP model modifies the relevant thermodynamic quantities and, in turn, the SGWB spectrum. Another issue raised in the GUP literature is the size of the deformation parameter. In principle, the dimensionless coupling $\beta_0$ could be of order unity, but the present constraints are still far too weak to fix its value. Depending on the observable and assumed GUP ansatz, laboratory experiments~\cite{Das:2008kaa,Scardigli:2014qka}, GWs observations~\cite{Feng:2016tyt} and cosmological probes such as big bang nucleosynthesis (BBN) can allow extremely large values of $|\beta_0|$, in some cases as high as $|\beta_0|\lesssim 10^{84}$–$10^{90}$ in the same higher-order GUP framework considered  here~\cite{Luo:2023rhk,Luo:2024vdd}. This very wide allowed interval indicates that it is currently neither possible nor meaningful to identify a unique “true” value of $\beta_0$. In the present analysis $\beta_0$ is therefore treated as a free, dimensionless parameter. Throughout the computations Planck units $\hbar = c = k_B = m_p = \ell_p =  1$ are adopted, so that $\beta$ and $\beta_0$ coincide numerically, and representative benchmark values of $\beta_0$ are employed to illustrate the qualitative impact of the higher-order GUP on the thermodynamics and on the resulting SGWB signals. For natural values $\beta_0=\mathcal{O}\left(1\right)$ the corrections at QCD temperatures are extremely small and well below the sensitivity of current detectors, lie well below the sensitivity of current detectors. Therefore, this analysis should be interpreted primarily as a phenomenologically sensitive study, and a constraint framework $\beta_0$ should be observed in the future for first-order SGWB on the QCD scale.

The main goal of this paper is to investigate how the Du–Long higher-order GUP affects the entropy of a photon gas and how these thermodynamic modifications propagate into the SGWB spectrum generated by a first-order phase transition at the QCD scale. The structure of the paper is as follows. Section~\ref{SECTII} derives the GUP-corrected entropy of a photon gas and the corresponding deformation of the phase-space measure within the Du–Long GUP model. Section~\ref{SECTIII} presents the construction of the effective equation of state employed in this work, by matching the ideal-gas thermodynamics to recent lattice QCD results around the QCD transition. In Section~\ref{SECTV} the influence of the higher-order GUP on the amplitude and frequency characteristics of the SGWB is investigated, with particular emphasis on the differences between positive and negative values of $\beta_0$ and on the physical consistency of the resulting thermodynamics. Finally, Section~\ref{SECTIV} summarizes the main findings and discusses possible extensions of the present analysis.

\section{The higher-order GUP corrected entropy and photons gas}
\label{SECTII}
In the framework of the new higher–order GUP~(\ref{eq1}), the canonical commutation relation between position and momentum operators is modified to include a position-dependent correction of the form
\begin{align}
\label{eq2}
\left[ {{x_i},{p_j}} \right] = i\hbar \frac{1}{{1 + \left( {{{16{\beta_0} } / {{x^2}}}} \right)}}{\delta _{ij}},
\end{align}
where $x$ is the characteristic length scale of the system. The modified commutation relation~(\ref{eq2}) implies the existence of a minimal measurable length or a maximal observable momentum, depending on the sign and magnitude of ${\beta_0}$. This deformation can be achieved through the representation of symmetry operators as follows
\begin{align}
\label{eq3}
{p_i} = \frac{1}{{1 + \left( {{{16{\beta_0} } / {{x^2}}}} \right)}}{p_{0i}},\quad {x_i} = {x_{0i}},
\end{align}
where $p_{0i}$ and $x_{0i}$ satisfy the usual Heisenberg algebra $\left[ {{x_{0i}},{p_{0j}}} \right] = i\hbar {\delta _{ij}}$. 

As a consequence of the modified commutation relation~(\ref{eq2}) and the operator representation~(\ref{eq3}), the canonical phase–space volume element is no longer invariant under time evolution. The transformation from the canonical variables $\left( {{x_{0i}},{p_{0i}}} \right)$ to the deformed variables $\left( {{x_i},{p_i}} \right)$ induces a non–trivial Jacobian in momentum space. According to Liouville's theorem, the number of quantum states in phase space must be conserved. This requirement leads to a deformation of the density of quantum states per unit spatial volume, which can be written in the form~\cite{Ali:2011ap}
\begin{align}
\label{eq4}
\frac{V}{{{{\left( {2\pi } \right)}^3}}}\int_0^\infty  {{\rm d}^3p}  \to \frac{V}{{{{\left( {2\pi } \right)}^3}}}\int_0^\infty  {{{\left( {1 + 16{\beta_0} {x^{ - 2}}} \right)}^3}{\rm d}^3p},
\end{align}
where the new higher–order GUP introduces an $x$–dependent factor in the momentum–space measure.

In order to investigate the thermodynamic consequences of the modified phase-space measure, it is convenient to consider a gas of massless bosons, in particular photons. In the early universe, the energy content was dominated by  radiation, with photons constituting a significant component of the thermal plasma. As a result of its zero mass and bosonic nature, the photon follows a Bose-Einstein distribution, providing an analytically tractable system for calculating partition functions and entropy. These thermodynamic quantities, once modified by the higher-order GUP, play a crucial role in determining the redshift of the SGWB generated during a QCD-scale first-order phase transition. In applying the deformed measure in eq.~(\ref{eq4}) to this cosmological setting, thermal equilibrium during the radiation-dominated epoch is assumed. Under such conditions, the spatial localization scale of relativistic particles is naturally characterized by the thermal de Broglie wavelength, $\lambda_{\mathrm{th}} \sim 1/T$ in natural units, which also sets the mean inter-particle spacing. It is therefore natural to identify the position uncertainty in the GUP with this thermal length scale, $\Delta x \sim T^{-1}$~\cite{KolbTurner1990}, and to interpret $x$ in eq.~(\ref{eq4}) as this coarse-grained uncertainty. This identification does not introduce a new fundamental uncertainty relation, but rather provides an effective description in a homogeneous radiation bath in thermal equilibrium. With this effective correspondence, the position-dependent correction factor in eq.~(\ref{eq4}) becomes a temperature-dependent one, and the number of quantum states per unit momentum-space volume is modified to
\begin{align}
\label{eq5}
\frac{V}{{{{\left( {2\pi } \right)}^3}}}\int_0^\infty  {{{\rm{d}}^3}p}  \to \frac{V}{{{{\left( {2\pi } \right)}^3}}}\int_0^\infty  {{{\left( {1 + 16{\beta_0} {T^2}} \right)}^3}{{\rm{d}}^3}p}.
\end{align}
According to eq.~(\ref{eq5}), the modified partition function per unit volume is
\begin{align}
\label{eq6}
\ln Z &= - \frac{{{g_\pi }}}{{2{\pi ^2}}}\int_0^\infty  {\ln \left[ {1 - \exp \left( { - \frac{p}{T}} \right)} \right]} {\left( {1 + 16{\beta_0} {T^2}} \right)^3}{p^2}{\rm{d}}p
\nonumber \\
&= \frac{{{g_\pi }{\pi ^2}}}{{90}}{T^3}{\left( {1 + 16 {\beta_0} {T^2}} \right)^3},
\end{align}
where ${g_\pi }$ denotes the number of internal degrees of freedom. In writing eq.~(\ref{eq6}), the deformation factor in eq.~(\ref{eq5}) simply multiplies the standard Bose--Einstein integrand in momentum space, while the remaining momentum integral is identical to that of an ordinary relativistic photon gas. The partition function~(\ref{eq6}) therefore encodes the effect of the higher-order GUP on the thermodynamics of the photon gas through a temperature-dependent correction factor.  From the deformed phase space and statistical distribution, the modified entropy of photon gas is given by
\begin{align}
\label{eq7}
S_{{\rm{GUP}}} &= \frac{\partial }{{\partial T}}\left( {T\ln Z} \right)= \frac{2}{{45}}{\pi ^2}{g_\pi }{T^3}\left( {1 + 72 {\beta_0} {T^2}+ 1536 {\beta_0}^2 {T^4} + 10240 {\beta_0}^3 {T^6}} \right).
\end{align}
When ignoring the effects of QG (i.e., ${\beta_0}  \to 0$), the modified entropy reduces smoothly to the standard expression ${S_0} = {{2{\pi ^2}{g_\pi }{T^3}} \mathord{\left/ {\vphantom {{2{\pi ^2}{g_\pi }{T^3}} {45}}} \right. \kern-\nulldelimiterspace} {45}}$ \cite{Kolb:1990vq}, ensuring consistency with classical thermodynamics. However, it is crucial to note that a negative deformation parameter
introduces an unphysical scenario: the thermodynamic quantities exhibit a maximum temperature $T_{\max} = m_p / \bigl(4\sqrt{|{\beta_0}|}\bigr)$, beyond which the entropy and pressure vanish. This clearly signals an internal inconsistency and unphysical behavior for negative ${\beta_0}$. In the subsequent analysis (see figure~\ref{fig1-b}), this inconsistency is explicitly demonstrated, further confirming that only the  positive-${\beta_0}$ scenario is physically viable.

\section{Impact of higher-order GUP modified entropy on the SGWB spectrum}
\label{SECTIII}

In investigating the influence of the GUP on the generation and evolution of the SGWB, a foundational task is to establish a thermodynamic description of the Universe that is consistent with the GUP-corrected phase-space measure. The introduction of a minimal measurable length or a maximal momentum, as predicted by GUP, modifies the phase-space volume (e.g., eq.~(\ref{eq4})), which in turn affects the statistical mechanics of relativistic radiation fields. In the framework of standard cosmology, the comoving entropy of the radiation fluid, dominated by photons, is given by $\mathbb{S} \sim a^3 S_0 \sim a^3\bigl(2\pi^2 g_s T^3/45\bigr)$, where $a$ is the scale factor and $g_s$ is the effective number of relativistic degrees of freedom. This quantity plays a central role in determining the dynamics of cosmic expansion. By contrast, in the presence of GUP-corrected minimal length or maximal momentum scales, both the density of quantum states and the occupation number are modified, leading to systematic changes in the statistical-mechanical quantities. In refs.~\cite{Moussa:2021qlz,Moussa:2021gxb,Khodadi:2018scn}, it have proposed GUP-corrected entropy expressions by modifying the density of states and typically retain only leading-order corrections (for example, terms linear or quadratic in temperature), an approximation that is justified when the GUP parameter is very small at accessible energy scales. Nevertheless, incorporating higher-order contributions can improve the consistency with fully non-perturbative GUP frameworks and may in principle reveal subtle cumulative QG effects, particularly in scenarios involving extremely high temperatures such as those near a QCD-scale first-order phase transition.

\begin{figure*}[htbp]
\centering
\subfigure[]{
\includegraphics[scale=0.73]{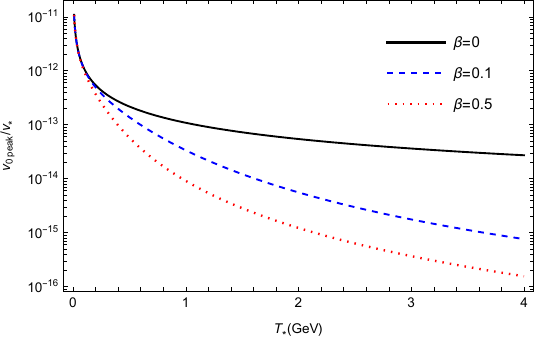}
\label{fig1-a}
}
\quad
\subfigure[]{
\includegraphics[scale=0.76]{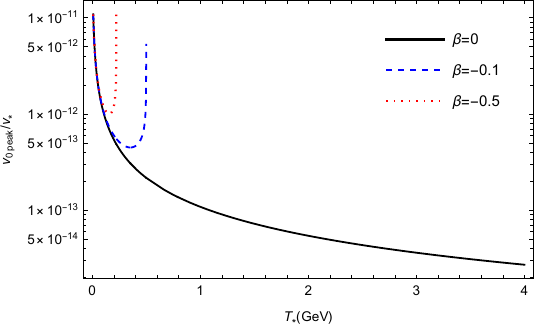}
\label{fig1-b}
}
\caption{The ratio ${\nu_{0{\rm peak}}}/{\nu_*}$ as a function of the transition temperature $T_*$ (in GeV) for different values of the dimensionless GUP parameter ${\beta_0}$. The present temperature is fixed at $T_0 = 2.725~{\rm K} = 2.348\times10^{-13}~{\rm GeV}$, with $g_s \left( {{T_0}} \right)=3.4$ and $g_s \left( {{T_*}} \right)=35$.  (a) Positive values of ${\beta_0}$. (b) Negative values of ${\beta_0}$.}
\label{fig1}
\end{figure*}

Now, applying the GUP corrected entropy~(\ref{eq7}), the relevant entropy density can be expressed as
\begin{align}
\label{eq8}
{\mathbb{S}_{{\rm{GUP}}}}  = {a^3}{g_s}\frac{2}{{45}}{\pi ^2}{T^3} \left( {1 + 72 {\beta_0} {T^2} + 1536{\beta_0}^2{T^4} + 10240{\beta_0}^3{T^6}} \right),
\end{align}
where $a$, $g_s$, and $T$ are functions of time $t$. Then, according to the adiabatic condition ${\dot{\mathbb{S}}/\mathbb{S} = 0}$ and using the relation ${\dot g_s} = \frac{{{\text{d}}{g_s}}}{{{\text{d}}T}}\frac{{{\text{d}}T}}{{{\text{d}}t}}$, which implies that the total entropy remains conserved during the expansion of the universe, the time variation of the universe temperature in the presence of GUP corrections as
\begin{align}
\label{eq9}
\frac{{{\rm{d}}T}}{{{\rm{d}}t}} =  - HT{\Delta}{\left( {T,g_s,{\beta_0} } \right)^{ - 1}},
\end{align}
with
\begin{align}
\label{eq10}
{\Delta}\left( {T,g_s,{\beta_0} } \right) = 1 + \frac{T}{{3{g_s}}}\frac{{{\rm{d}}{g_s}}}{{{\rm{d}}T}} + \frac{{16{\beta_0} {T^2}\left( {3+80{\beta_0} {T^2}} \right)}}{{1 + 56{\beta_0} {T^2} + 640{{\beta_0} ^2}{T^4}}}.
\end{align}
For ${\beta_0} = 0$, the original  case $\Delta  = 1 + \left( {{T \mathord{\left/ {\vphantom {T {3{g_s}}}} \right. \kern-\nulldelimiterspace} {3{g_s}}}} \right)\left( {{{{\rm{d}}{g_s}} \mathord{\left/ {\vphantom {{{\rm{d}}{g_s}} {{\rm{d}}T}}} \right. \kern-\nulldelimiterspace} {{\rm{d}}T}}} \right)$ is recovered. Due to the Hubble parameter, one has
\begin{align}
\label{eq11}
\frac{{{a_*}}}{{{a_0}}} = \exp \left( {\int_{{T_*}}^{{T_0}} {{T^{ - 1}}\Delta {\rm{d}}T} } \right),
\end{align}
where the indices ``*" and ``0" denote the corresponding physical parameters evaluated at the time of the phase transition and the present era, respectively. Using the relationship between the scale factor and redshift ${{{\nu _{0{\rm{peak}}}}} \mathord{\left/ {\vphantom {{{\nu _{0{\rm{peak}}}}} {{\nu _*}}}} \right.
 \kern-\nulldelimiterspace} {{\nu _*}}} = {{{a_*}} \mathord{\left/
 {\vphantom {{{a_*}} {{a_0}}}} \right.
 \kern-\nulldelimiterspace} {{a_0}}}$, the redshift corresponding to the peak frequency of SGWB relative to its present value can be formulated as follows
\begin{align}
\label{eq12}
\frac{{{\nu _{0{\rm{peak}}}}}}{{{\nu _*}}} = \frac{{{a_*}}}{{{a_0}}} = \frac{{{T_0}}}{{{T_*}}}{\left[ {\frac{{{g_s}\left( {{T_0}} \right)}}{{{g_s}\left( {{T_*}} \right)}}} \right]^{\frac{1}{3}}}{\left( {\frac{{1 + 16{\beta_0} T_0^2}}{{1 + 16{\beta_0} T_*^2}}} \right)^{\frac{2}{3}}}{\left( {\frac{{1 + 40{\beta_0} T_0^2}}{{1 + 40{\beta_0} T_*^2}}} \right)^{\frac{1}{3}}}.
\end{align}
To examine the role of the GUP effect in modifying the ratio of the peak frequency of the present SGWB to the frequency at the QCD phase transition epoch, we plot figure~\ref{fig1}.

Figure~\ref{fig1} shows how the ratio of the peak frequency of the
present SGWB to the frequency at the QCD phase transition epoch,
${\nu_{0{\rm peak}}}/{\nu_*}$, varies with the QCD phase transition
temperature $T_*$ for different values of the GUP parameter ${\beta_0}$. Here ${T_0} = 2.725K = 2.348 \times {10^{-13}}{\text{GeV}}$, $g_s \left( {{T_0}} \right)=3.4$ and $g_s \left( {{T_*}} \right)=35$~\cite{Moussa:2021qlz,Moussa:2021gxb}. Notably,  $T_*$ denotes the physical transition temperature (in GeV), while the GUP corrections are expressed in terms of the dimensionless parameter ${\beta_0}$ introduced in the previous section. For numerical evaluation, the temperature entering the GUP correction factor is the dimensionless variable $T{\text{ =  }}{{{T_*}} \mathord{\left/ {\vphantom {{{T_*}} {{m_p}}}} \right. \kern-\nulldelimiterspace} {{m_p}}}$ (with $m_p=1$ in Planck units), whereas the horizontal axis in Fig.~\ref{fig1} is labeled with the corresponding physical temperature $T_*$.

Without the effects of GUP  (${\beta_0} = 0$), the frequency ratio demonstrates a stable decreasing trend. With increasing positive ${\beta_0}$ values, the frequency ratio decreases significantly, indicating a strong suppression effect of positive GUP parameters on the current peak frequency of SGWB. This suppression effect becomes more pronounced as the transition temperature increases. Figure~\ref{fig1-b} presents the scenario with negative values of ${\beta_0}$, revealing a qualitatively opposite behavior. As the absolute value of negative ${\beta_0}$ increases, the frequency ratio grows rapidly with increasing transition temperature and eventually diverges at a certain high temperature. This divergence explicitly demonstrates the theoretical expectation discussed above: a negative deformation
parameter leads to unphysical thermodynamic behavior characterized by a maximum temperature, beyond which the entropy and pressure vanish. Such behavior signals a fundamental internal inconsistency of the negative ${\beta_0}$ scenario, particularly at temperatures approaching the Planck scale. Therefore, in the following analysis only the case of positive ${\beta_0}$ will be considered.

We now turn to the evolution of the SGWB energy density. Since GWs decouple from the thermal plasma shortly after their generation, their energy density satisfies the Boltzmann equation ${{{\rm{d}}\left( {{\rho _{{\rm{gw}}}}{a^4}} \right)} \mathord{\left/ {\vphantom {{{\rm{d}}\left( {{\rho _{{\rm{gw}}}}{a^4}} \right)} {{\rm{d}}t}}} \right. \kern-\nulldelimiterspace} {{\rm{d}}t}} = 0$. This conservation law allows one to follow the redshift of the SGWB energy density from the phase-transition epoch to the present era. The energy density of SGWB at the present time is then given by
\begin{align}
\label{eq13}
{\rho _{{\rm{gw}}}}\left( {{T_0}} \right)  = {\rho _{{\rm{gw}}}}\left( {{T_*}} \right){\left( {\frac{{{a_*}}}{{{a_0}}}} \right)^4}= {\rho _{{\rm{gw}}}}\left( {{T_*}} \right)\exp \left( {\int_{{T_*}}^{{T_0}} {4{T^{ - 1}}} \Delta {\rm{d}}T} \right).
\end{align}

Next, based on eq.~(\ref{eq13}) and the definitions of the SGWB energy density parameters at the phase transition temperature $T_*$ and the present temperature $T_0$, given by ${\Omega _{{\rm{gw_*}}}} = {{{\rho _{{\rm{gw}}}}\left( {{T_*}} \right)} \mathord{\left/
 {\vphantom {{{\rho _{{\rm{gw}}}}\left( {{T_*}} \right)} {{\rho _{cr}}\left( {{T_*}} \right)}}} \right.
 \kern-\nulldelimiterspace} {{\rho _{cr}}\left( {{T_*}} \right)}}$ and ${\Omega _{{\rm{gw}}}} = {{{\rho _{{\rm{gw}}}}\left( {{T_0}} \right)} \mathord{\left/
 {\vphantom {{{\rho _{{\rm{gw}}}}\left( {{T_0}} \right)} {{\rho _{\rm{cr}}}\left( {{T_0}} \right)}}} \right.
 \kern-\nulldelimiterspace} {{\rho _{\rm{cr}}}\left( {{T_0}} \right)}}$, the expression for the SGWB energy density at present time reads
\begin{align}
\label{eq14L}
{\Omega _{{\rm{gw}}}} = {\Omega _{{\rm{gw_*}}}}{\left( {\frac{{{H_*}}}{{{H_0}}}} \right)^2}\exp \left( {\int_{{T_*}}^{{T_0}} {4{T^{ - 1}}} \Delta {\rm{d}}T} \right)
\end{align}
where
\begin{align}
\label{eq14}
{\left( {\frac{{{H_*}}}{{{H_0}}}} \right)^2} = \frac{{{\rho _{{\rm{cr}}}}\left( {{T_*}} \right)}}{{{\rho _{{\rm{cr}}}}\left( {{T_0}} \right)}}.
\end{align}

For the purpose of gaining the ratio of Hubble parameter during the cosmic phase transition to its present-day value, we are supposed to leverage the continuity equation $\dot{\rho}_t = -3H \rho_t \left(1 + \frac{P_t}{\rho_t}\right)$, where $P_t$ represents the total pressure density of the universe and $\rho_t$ is the total energy density of the universe, respectively. By expressing the continuity equation in terms of temperature and substituting eq.~(\ref{eq9}), one gets
\begin{align}
\label{eqa13}
\frac{{{\rm{d}}{\rho _t}}}{{{\rho _t}}} = \frac{3}{T}\left( {1 + {\omega _{{\rm{eff }}}}} \right)\Delta {\rm{d}}T,
\end{align}
with the effective EOS parameter ${\omega _{{\rm{eff}}}}$ $= {{{P_t}} \mathord{\left/ {\vphantom {{{P_t}} {{\rho _t}}}} \right. \kern-\nulldelimiterspace} {{\rho _t}}}$. By integrating eq.~(\ref{eq9}) from the early radiation-dominated era (characterized by a temperature $T_r = 10^4  \mathrm{GeV}$) to the phase transition temperature $T_*$, the critical energy density of radiation during the phase transition is given by
\begin{align}
\label{eqa14}
{\rho _{{\rm{cr}}}}\left( {{T_*}} \right) = {\rho _r}\left( {{T_r}} \right)\exp \left[ {\int_{{T_r}}^{{T_*}} {\frac{3}{T}} \left( {1 + {\omega _{{\rm{eff}}}}} \right)\Delta {\rm{d}}T} \right].
\end{align}
Then, by substituting $\rho_{\rm{cr}}(T_*)$ from eq.~(\ref{eqa14}) into eq.~(\ref{eq14}), one obtains
\begin{align}
\label{eqa15}
{\left( {\frac{{{H_*}}}{{{H_0}}}} \right)^2} \! = \!{\Omega _{r0}}\frac{{{\rho _r}\left( {{T_r}} \right)}}{{{\rho _r}\left( {{T_0}} \right)}}\exp \left[ {\int_{{T_r}}^{{T_*}} {\frac{3}{T}} \left( {1 + {\omega _{{\rm{eff }}}}} \right)\Delta {\rm{d}}T} \right],
\end{align}
where the current fractional energy density of radiation is $\Omega_{r 0}=\rho_{r}\left(T_{0}\right) / \rho_{c r}\left(T_{0}\right) \simeq 8.5 \times 10^{-5}$. Moreover, by applying the Boltzmann equation, it is found that  $\rho_r(T_r)/\rho_r(T_0) \simeq (a_0/a_r)^4$ \cite{Moussa:2021gxb}. Therefore, eq.~(\ref{eqa15})  can be rewritten as
\begin{align}
\label{eqa16}
{\left( {\frac{{{H_*}}}{{{H_0}}}} \right)^2} & = {\Omega _{r0}}\exp \left( {\int_{{T_0}}^{{T_r}} {\frac{4}{T}} \Delta {\rm{d}}T} \right)\exp \left[ {\int_{{T_r}}^{{T_*}} {\frac{3}{T}} \left( {1 + {\omega _{{\rm{eff }}}}} \right)\Delta {\rm{d}}T} \right].
\end{align}
Finally, by using eq.~(\ref{eqa16}) and eq.~(\ref{eq14L}), the GWs spectrum observed today can be expressed as
\begin{align}
\label{eqa17}
{\Omega _{{\rm{gw}}}} & = {\Omega _{r0}}{\Omega _{{\rm{g}}{{\rm{w}}_*}}}\exp \left( {\int_{{T_*}}^{{T_r}} {\frac{4}{T}} \Delta {\rm{d}}T} \right) \exp \left[ {\int_{{T_r}}^{{T_*}} {\frac{3}{T}} \left( {1 + {\omega _{{\rm{eff}}}}} \right)\Delta {\rm{d}}T} \right].
\end{align}
The above equation gives the present-day energy-density spectrum of the SGWB, relating the spectrum at the phase-transition epoch, $\Omega_{\rm gw,*}$, to its observable form today, $\Omega_{\rm gw,0}$. The factor $\Omega_{r0}$ represents the current radiation energy density, while the exponential terms encode the effect of the intervening expansion history. More importantly, the integrand contains the function $\Delta(T,g_s,\beta_0)$, which parametrizes the modifications induced by the Du-Long higher-order GUP. The GUP corrections modify the standard thermodynamic relations, thereby influencing the evolution of the Hubble parameter and the redshift of SGWB.

\section{Influence of non-ideal equation of state on SGWB during the QCD phase transition}
\label{SECTIIII}
In this section, the impact of the effective equation of state (EOS) on the spectral characteristics of the SGWB is examined, with particular attention to the role of non-ideal QCD interactions and trace anomaly effects. These effects become especially important in the QCD transition region, where strong interactions among quarks and gluons lead to substantial deviations from the ideal relativistic-gas approximation commonly used in cosmological models. In addition, the influence of the new higher-order GUP corrections on the EOS and on the subsequent evolution of the SGWB spectrum is investigated, thereby assessing the relevance of quantum-gravity effects in this cosmological epoch.

For an ultra-relativistic ideal gas composed of non-interacting particles, the effective EOS parameter is given by ${{\omega _{{\rm{eff}}}} = {1 \mathord{\left/ {\vphantom {1 3}} \right. \kern-\nulldelimiterspace} 3}}$ \cite{Anand:2017kar}. Under this condition, eq.~(\ref{eqa16}) and eq.~(\ref{eqa17}) can be recast into the following analytical expressions
\begin{align}
\label{eq15}
{\left( {\frac{{{H_*}}}{{{H_0}}}} \right)^2} & = {\Omega _{r0}}{\left( {\frac{{{T_*}}}{{{T_0}}}} \right)^4}{\left[ {\frac{{{g_s}\left( {{T_*}} \right)}}{{{g_s} \left( {{T_0}} \right)}}} \right]^{\frac{4}{3}}} {\left( {\frac{{1 + 16{\beta_0} T_*^2}}{{1 + 16{\beta_0} T_0^2}}} \right)^{\frac{8}{3}}}{\left( {\frac{{1 + 40{\beta_0} T_*^2}}{{1 + 40{\beta_0} T_0^2}}} \right)^{\frac{4}{3}}},
\end{align}
\begin{align}
\label{eq16}
{\Omega _{{\rm{gw}}}} = {\Omega _{r0}}{\Omega _{{\rm{gw}}}},
\end{align}
Notably, the present-day SGWB spectrum described by eq.~(\ref{eq16}) reverts to its standard form, indicating that the GUP parameter ${\beta_0}$ ceases to influence the observed spectral features. However, this idealized scenario diverges from reality when QCD interactions are incorporated, as they induce deviations from $\omega_{\rm{eff}} = 1/3$. These deviations can substantially modify the SGWB spectrum relative to the non-interacting case, highlighting the importance of incorporating a refined EOS during the QCD epoch \cite{HotQCD:2014kol}.

To incorporate the effects of QCD interactions, we employ results from modern lattice QCD calculations using $N_f = 2+1$ flavors, which reliably cover the temperature range from $0.1~\rm{GeV}$ to $0.4~\rm{GeV}$~\cite{HotQCD:2014kol} \footnote{Although these lattice results describe a crossover transition at vanishing baryon chemical potential, in the present work they are used as input for a phenomenological model of a hypothetical first-order phase transition at the same QCD temperature scale, in order to compute the associated SGWB background.}. These calculations enable a realistic description of the strongly interacting plasma. Accordingly, the QCD equation of state is constructed by adopting a parametrization of the pressure that reproduces the lattice results, effectively encoding the contributions from the strong interactions among $u$, $d$ and $s$ quarks and gluons. The parametrized expression for the QCD pressure reads 
\begin{align}
\label{eq15L}
F\left( T \right) = \frac{P}{{{T^4}}} \frac{{1 + \tanh \left[ {{c_\tau }\left( {\tau  - {\tau _0}} \right)} \right]}}{2}\frac{{{p_i} + \frac{{{a_n}}}{\tau } + \frac{{{b_n}}}{{{\tau ^2}}} + \frac{{{c_n}}}{{{\tau ^4}}}}}{{1 + \frac{{{a_d}}}{\tau } + \frac{{{b_d}}}{{{\tau ^2}}} + \frac{{{c_d}}}{{{\tau ^4}}}}}.
\end{align}
Here, the dimensionless temperature variable is defined as $\tau = T/T_c$, where $\tau  = {T \mathord{\left/ {\vphantom {T {{T_c}}}} \right. \kern-\nulldelimiterspace} {{T_c}}}$ is taken as a representative value of the QCD transition temperature.  For a system of three massless quark flavors, the ideal gas limit of the normalized pressure is given by ${p_i} = {{19{\pi ^2}} \mathord{\left/  {\vphantom {{19{\pi ^2}} {36}}} \right.  \kern-\nulldelimiterspace} {36}}$. At temperatures above $100~\rm{MeV}$, and the fitting parameters are specified as follows:
\begin{align}
\label{eq16L}
&c_\tau \!= \! 3.876,a_n \! = \! -8.7704, b_n\! = \!3.9200, c_n = 0.3419,
\nonumber \\
&\tau_0 \!= \!0.9761,a_d\! = \!-1.2600,  b_d \!=\! 0.8425,  c_d \!= \!-0.0475.
\end{align}
Notably, in the QCD transition region the EOS deviates significantly from ideal-gas behavior because of strong interactions and the breaking of approximate conformal invariance, so that the pressure alone is not sufficient to fully characterize the thermodynamic properties of the system. To quantify these deviations, the trace-anomaly (interaction-measure) relation is employed~\cite{Cheng:2007jq}, which relates the pressure and the energy density through the non-vanishing trace of the QCD energy–momentum tensor:
\begin{align}
\label{eq17L}
\frac{\rho - 3P}{T^4} = T \frac{\mathrm{d}F(T)}{\mathrm{d}T} = T \frac{\mathrm{d}}{\mathrm{d}T} \left( \frac{P}{T^4} \right),
\end{align}
Accordingly, when the effects of QCD are taken into account, the effective EOS parameter that incorporates the trace anomaly can be expressed as
\begin{align}
\label{eq18L}
\omega_{\mathrm{eff}} = \left[ \frac{T}{F\left(T\right)} \frac{\mathrm{d}F \left(T\right)}{\mathrm{d}T} + 3 \right]^{-1}.
\end{align}

\begin{figure}
\centering
\includegraphics[width=0.59\linewidth]{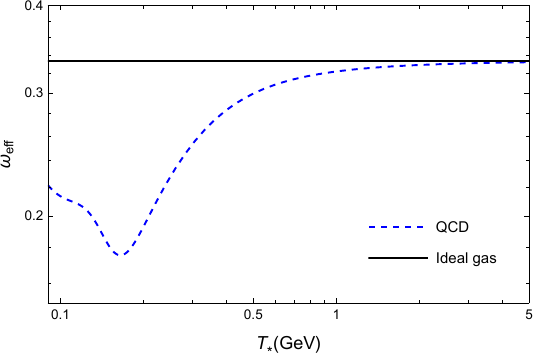}
\caption{The relationship between the  effective EOS $\omega_{\rm{eff}}$ and transition temperature $T_*$.}
\label{fig2}
\end{figure}

Based on eq.~(\ref{eq18L}), the relationship between the effective EOS $\omega_{\rm{eff}}$ and the transition temperature $ T_*$ is illustrated in figure~\ref{fig2}. It can be seen that, at high temperatures ($T_* \ge 5~\rm{GeV}$) asymptotically approaches the ideal gas limit of ${1 \mathord{\left/{\vphantom {1 3}} \right. \kern-\nulldelimiterspace} 3}$, indicating negligible interaction effects. However, in the lower temperature range (${T_*} \sim 0.1 - 0.3~\rm{GeV}$), which corresponds to the QCD transition region, $\omega_{\rm{eff}}$ deviates significantly from the ideal value due to the strong interactions among quarks and gluons. This behavior underscores the necessity of incorporating realistic QCD effects into the SGWB analysis.

Then, by substituting the effective EOS parameter incorporating the trace anomaly~(\ref{eq18L}) into eq.~(\ref{eqa17}), the Hubble parameter ratio between the transition epoch and the present epoch can be expressed as
\begin{align}
\label{eq17}
{\left( {\frac{{{H_*}}}{{{H_0}}}} \right)^2} & = {\Omega _{r0}}{\left( {\frac{{{T_r}}}{{{T_0}}}} \right)^4}{\left[ {\frac{{{g_s}\left( {{T_r}} \right)}}{{{g_s}\left( {{T_0}} \right)}}} \right]^{\frac{4}{3}}}{\left( {\frac{{1 + 16{\beta_0} T_r^2}}{{1 + 16{\beta_0} T_0^2}}} \right)^{\frac{8}{3}}}
\times {\left( {\frac{{1 + 40{\beta_0} T_r^2}}{{1 + 40{\beta_0} T_0^2}}} \right)^{\frac{4}{3}}}\frac{{{g_s}{{\left( {{T_*}} \right)}^{1 + \omega \left( {{T_*}} \right)}}}}{{{g_s}{{\left( {{T_r}} \right)}^{1 + \omega \left( {{T_r}} \right)}}}}
\nonumber \\
&\times  \exp \left\{ {\int_{{T_r}}^{{T_*}} {\frac{{3\left( {1 + \omega } \right)}}{T}\left[ {1 + \frac{{16{\beta_0} {T^2}\left( {3 + 80{\beta_0} {T^2}} \right)}}{{1 + 56{\beta_0} {T^2} + 640{{\beta_0} ^2}{T^4}}}} \right]{\rm{d}}T} } \right\},
\end{align}

\begin{figure}[htbp]
\centering
\subfigure[]{
\includegraphics[scale=0.79]{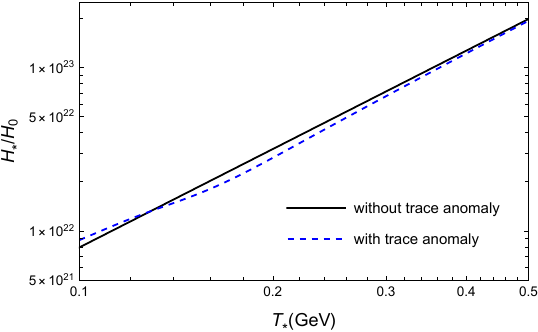}
\label{fig3-a}
}
\quad
\subfigure[]{
\includegraphics[scale=0.79]{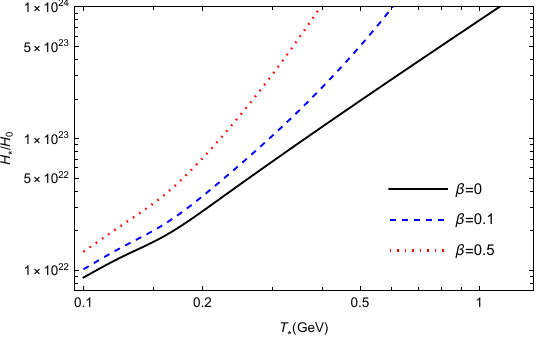}
\label{fig3-b}
}
\caption{The relationship between the  ${{{H_*}} \mathord{\left/
 {\vphantom {{{H_*}} {{H_0}}}} \right.
 \kern-\nulldelimiterspace} {{H_0}}}$ and transition temperature $T_*$. We set $T_r=10^4~\rm{GeV}$ and ${g_s}\left( {{T_r}} \right) = 106$. (a) Without the effect of GUP.
(b) With different values of GUP parameter.}
\label{fig3}
\end{figure}
From figure~\ref{fig3}, one can see the variation of the Hubble parameter ratio ${{{{H_*}} \mathord{\left/{\vphantom {{{H_*}} {{H_0}}}} \right.\kern-\nulldelimiterspace} {{H_0}}}}$ as a function of the QCD transition temperature $T_*$ under different conditions, with $T_r=10^4~\rm{GeV}$ and ${g_s}\left( {{T_r}} \right) = 106$~\cite{Moussa:2021qlz,Moussa:2021gxb}, according to eq.~(\ref{eq17}). Figure~\ref{fig3-a} compares the scenarios with (blue dashed curve) and without (black solid curve) the inclusion of the QCD trace anomaly effect. In the lower temperature range (${T_*} \sim 0.1 - 0.15~\rm{GeV}$), the blue dashed curve is notably above the black solid curve, indicating that the trace anomaly significantly enhances the Hubble parameter ratio. However, as the transition temperature increases to the intermediate range (${T_*} \sim 0.15 - 0.5~\rm{GeV}$), the trend reverses, and the black solid line exceeds the blue dashed line, indicating a reduced impact of the trace anomaly effect compared to the ideal gas case. When the temperature is higher than $0.5~\rm{GeV}$, the two curves gradually converge, showing that QCD interactions become negligible, and both cases approach the ideal relativistic gas limit. These results highlight the temperature-dependent significance of the trace anomaly and the necessity for accurate QCD thermodynamics when analyzing the cosmological evolution of the SGWB. Figure~\ref{fig3-b} illustrates the influence of positive Du--Long higher-order GUP parameters, with $\beta_0 = 0.1$ and $\beta_0 = 0.5$, on the ratio $H_/H_0$. Increasing $\beta_0$ leads to a noticeable enhancement of $H_/H_0$, especially at higher transition temperatures, reflecting the fact that stronger GUP corrections accelerate the expansion rate of the early Universe in this framework. These results highlight the combined impact of QCD thermodynamics and quantum-gravity corrections on the expansion history relevant for the SGWB.

Then, incorporating the influence of the trace anomaly along with eq.~(\ref{eqa17}), the ratio of the GW energy density parameter is given by
\begin{align}
\label{eq18}
\frac{{{\Omega _{{\rm{gw}}}}}}{{{\Omega _{{\rm{gw*}}}}}} & = {\Omega _{r0}}{\left( {\frac{{{T_r}}}{{{T_*}}}} \right)^4}{\left( {\frac{{{g_s}\left( {{T_r}} \right)}}{{{g_s}\left( {{T_*}} \right)}}} \right)^{\frac{4}{3}}} {\left( {\frac{{1 + 16{\beta_0} T_r^2}}{{1 + 16{\beta_0} T_*^2}}} \right)^{\frac{8}{3}}}{\left( {\frac{{1 + 40{\beta_0} T_r^2}}{{1 + 40{\beta_0} T_*^2}}} \right)^{\frac{4}{3}}} \frac{{{g_s}{{\left( {{T_*}} \right)}^{1 + \omega \left( {{T_*}} \right)}}}}{{{g_s}{{\left( {{T_r}} \right)}^{1 + \omega \left( {{T_r}} \right)}}}}
\nonumber \\
& \times \exp \left\{ {\int_{{T_r}}^{{T_*}} {\frac{{3\left( {1 + \omega } \right)}}{T}\left[ {1 + \frac{{16{\beta_0} {T^2}\left( {3 + 80{\beta_0} {T^2}} \right)}}{{1 + 56{\beta_0} {T^2} + 640{{\beta_0} ^2}{T^4}}}} \right]{\rm{d}}T} } \right\}.
\end{align}

\begin{figure}[htbp]
\centering
\includegraphics[width=0.6\linewidth]{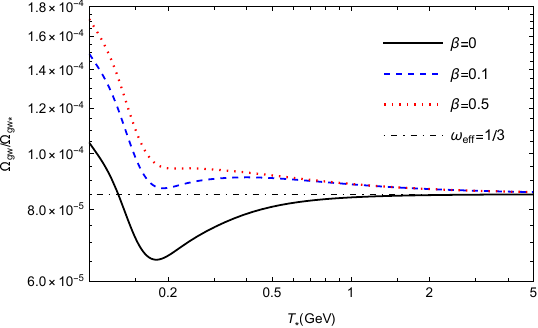}
\caption{The relationship between ${{{\Omega _{{\rm{gw}}}}} \mathord{\left/
 {\vphantom {{{\Omega _{{\rm{gw}}}}} {{\Omega _{{\rm{gw*}}}}}}} \right.
 \kern-\nulldelimiterspace} {{\Omega _{{\rm{gw*}}}}}}$ and  the function
of transition temperature $T_*$ with different values of GUP parameter ${\beta_0}$.}
\label{fig5}
\end{figure}

In figure~\ref{fig5}, we present the ratio of the GW energy density ${{{\Omega _{{\rm{gw}}}}} \mathord{\left/
 {\vphantom {{{\Omega _{{\rm{gw}}}}} {{\Omega _{{\rm{gw*}}}}}}} \right.
 \kern-\nulldelimiterspace} {{\Omega _{{\rm{gw*}}}}}}$ at present times to the energy density at the QCD phase transition epoch, as a function of the QCD phase transition temperature $T$ for different values of the dimensionless GUP parameter ${\beta_0}$. The three curves represent the cases with ${\beta_0}=0$ (black solid curve), ${\beta_0}=0.1$ (blue dashed curve), and ${\beta_0}=0.5$ (red dotted curve). A horizontal gray dotted line indicates the reference value obtained assuming an ideal ultra-relativistic gas with ${{\omega _{{\rm{eff}}}} = {1 \mathord{\left/ {\vphantom {1 3}} \right. \kern-\nulldelimiterspace} 3}}$. Increasing ${\beta_0}$ results in a significant increase in the GW energy density ratio, especially in the temperature range (${T_*} \sim 0.1 - 0.3~\rm{GeV}$), where QCD interactions are most prominent. This increase arises from the GUP-induced modifications to the thermodynamic quantities, which in turn affect the redshift behavior of the SGWB. At higher temperatures (${T_*} > 0.5~\rm{GeV}$), all curves converge to the gray reference line, reflecting the restoration of standard cosmological behavior, where QG and QCD corrections become negligible. These results demonstrate that the GUP effects, even at sub-Planckian temperatures, can leave observable imprints on the SGWB spectrum, providing a potential probe into QG phenomena.

\section{Modified QCD sources of SGWB}
\label{SECTV}
After examining how the QCD equation of state and QG effects influence the evolution of the SGWB, we now focus on its generation mechanisms originating from the QCD phase transition. Specifically, if the transition is first-order, it can become a significant source of SGWB production through three primary mechanisms: expanding bubble wall collisions~\cite{Kosowsky:1991ua,Kosowsky:1992rz,Huber:2008hg}, sound waves in the plasma~\cite{Hindmarsh:2013xza,Hindmarsh:2015qta}, and magnetohydrodynamic turbulence~\cite{Kamionkowski:1993fg,Binetruy:2012ze}. These processes collectively contribute to the present-day SGWB spectrum ${\Omega _{{\rm{gw}}_{*}}} = {\Omega _{{\rm{gw}}}}\left( {{T_*}} \right)$ as described by the following expression: 

Bubble wall collisions (BC)~\cite{Caprini:2007xq,Jinno:2017fby}
\begin{align}
\label{eq19}
\Omega _{{\rm{g}}{{\rm{w}}_{_*}}}^{{\rm{BC}}}\left( \nu  \right) = {\left( {\frac{{{H_*}}}{\alpha }} \right)^2}{\chi _{{\rm{BC}}}}\left( {\frac{{0.11{\mu ^3}}}{{0.42 + {\mu ^2}}}} \right){S_{{\rm{BC}}}}, \quad {\chi _{{\rm{BC}}}} = {\left( {\frac{{{\kappa _{{\rm{BC}}}}\epsilon}}{{1 + \epsilon}}} \right)^2}.
\end{align}

Sound waves (SW)~\cite{Hindmarsh:2017gnf}
\begin{align}
\label{eq20}
\Omega _{{\rm{g}}{{\rm{w}}_{_*}}}^{{\rm{SW}}}\left( \nu  \right) = {\left( {\frac{{{H_*}}}{\alpha }} \right)^2}{\chi _{{\rm{SW}}}}\mu {S_{{\rm{SW}}}}, \quad
{\chi _{{\rm{SW}}}} = {\left( {\frac{{{\kappa _{{\rm{SW}}}}}\epsilon}{{1 + \epsilon}}} \right)^2}.
\end{align}

Magnetohydrodynamic (MHD) turbulence~\cite{Caprini:2006jb,Gogoberidze:2007an,Caprini:2009yp}
\begin{align}
\label{eq21}
\Omega _{{\rm{g}}{{\rm{w}}_{_*}}}^{{\rm{MHD}}}\left( \nu  \right) = \left( {\frac{{{H_*}}}{\alpha }} \right){\chi _{{\rm{MHD}}}}{S_{{\rm{MHD}}}}, \quad
{\chi _{{\rm{MHD}}}} &= {\left( {\frac{{{\kappa _{{\rm{MHD}}}}}\epsilon}{{1 + \epsilon}}} \right)^{\frac{3}{2}}},
\end{align}
where $H_*$ denotes the Hubble parameter at the time of production of SGWB, while $\alpha$ characterizes the time of the phase transition, and $\mu$ represents the bubble wall velocity. The parameter $\epsilon$ quantifies the ratio of the vacuum energy released during the phase transition to the radiation energy density. Furthermore, the factors $\kappa_{\rm{BC}}$, $\kappa_{\rm{SW}}$ and $\kappa_{\rm{MHD}}$ describe the fractional allocation of the phase transition-latent heat to BC, SW, and MHD turbulence, respectively. The functions of the SGWB which are characterized from numerical fits as
\begin{align}
\label{eq22}
{S_{{\rm{BC}}}} & = \frac{{3.8{{\left( {{\nu  \mathord{\left/ {\vphantom {\nu  {{\nu _{{\rm{BC}}}}}}} \right. \kern-\nulldelimiterspace} {{\nu _{{\rm{BC}}}}}}} \right)}^{2.8}}}}{{1 + 2.8{{\left( {{\nu  \mathord{\left/
 {\vphantom {\nu  {{\nu _{{\rm{BC}}}}}}} \right. \kern-\nulldelimiterspace} {{\nu _{{\rm{BC}}}}}}} \right)}^{3.8}}}},
\\
{S_{{\rm{SW}}}} & = {{\left( {\frac{\nu}{{{\nu_{{\rm{SW}}}}}}} \right)}^3}{{\left[ {\frac{7}{{4 + 3{{\left( {{\nu \mathord{\left/
 {\vphantom {\nu {{\nu_{{\rm{SW}}}}}}} \right.
 \kern-\nulldelimiterspace} {{\nu_{{\rm{SW}}}}}}} \right)}^2}}}} \right]}^{3.5}},
 \\
{S_{{\rm{MHD}}}} & = \frac{{\mu {{\left( {\frac{\nu }{{{\nu _{{\rm{MHD}}}}}}} \right)}^3}}}{{{{\left( {1 + \frac{\nu }{{{\nu _{{\rm{MHD}}}}}}} \right)}^{\frac{{11}}{3}}}\left[ {1 + \frac{{8\pi \nu }}{{{H_*}}}{{\left( {\frac{{{a_*}}}{{{a_0}}}} \right)}^{ - 1}}} \right]}},
\end{align}
and the current peak frequencies  of the SGWB  generated by BC, WS, and MHD at the time of phase transition are given by
\begin{align}
\label{eq22+}
\nu_{{\rm{BC }}}  = \frac{0.62 \alpha}{1.8 - 0.1 \mu + \mu^2} \left( \frac{a_*}{a_0} \right), \quad
\nu_{{\rm{SW }}} = \frac{38 \alpha}{31 \mu} \left( \frac{a_*}{a_0} \right), \quad \nu_{{\rm{MHD }}} = \frac{7 \alpha}{4 \mu} \left( \frac{a_*}{a_0} \right).
\end{align}
The parameters $\epsilon$ and $\kappa$, which appear within $\chi$, significantly influence both the spectral peak location and amplitude of the SGWB signal. Due to their strong model dependence, however, a universally reliable analytical form for $\kappa$ has yet to be established. Following the assumptions adopted in Refs.~\cite{Anand:2017kar,Khodadi:2021ees}, we take ${{{\kappa _{{\text{BC}}}}\epsilon } \mathord{\left/
 {\vphantom {{{\kappa _{{\text{BC}}}}\epsilon } {\left( {1 + \epsilon } \right)}}} \right.
 \kern-\nulldelimiterspace} {\left( {1 + \epsilon } \right)}} = {{{\kappa _{{\text{SW}}}}\epsilon } \mathord{\left/
 {\vphantom {{{\kappa _{{\text{SW}}}}\epsilon } {\left( {1 + \epsilon } \right)}}} \right.
 \kern-\nulldelimiterspace} {\left( {1 + \epsilon } \right)}} = {{{\kappa _{{\text{MHD}}}}\epsilon } \mathord{\left/
 {\vphantom {{{\kappa _{{\text{MHD}}}}\epsilon } {\left( {1 + \epsilon } \right)}}} \right.
 \kern-\nulldelimiterspace} {\left( {1 + \epsilon } \right)}} = 0.05$, adopt the relation $\alpha = nH_{*}$ with $n = 5$, and fix the bubble wall velocity as $\mu = 0.7$. Under these assumptions, the Hubble parameter at the time of the phase transition is given by

\begin{align}
\label{eq23}
{H_*} = \sqrt {\frac{{8\pi }}{{3m_p^2}}\rho \left( {{T_*}} \right)},
\end{align}
with the Planck mass $m_p$ and the energy density at transition temperature $
\rho\left(T_{*}\right) = T_{*}^5 \left[ \mathrm{d}F \left(T_{*} \right)/\mathrm{d}T_{*} \right] + 3T_{*}^4 F \left(T_{*}\right)$. It is well established that QCD phase transitions occur at temperatures ranging from a few hundred MeV, though the precise value depends on the details of QCD matter content. For simplicity, we assume that gravitational waves are generated instantaneously at the time of the phase transition, and thus set the transition temperature to the critical temperature $T_c=T_*=0.145~\rm{GeV}$. This assumption allows us to directly relate the early-universe thermodynamics to present-day observable quantities \footnote{In this work, our primary objective is to investigate the influence of higher-order GUP corrections on the SGWB generated during a QCD phase transition. To focus on the effects of QG, we adopt a commonly used assumption of a sharp first-order QCD phase transition at a fixed temperature $T_*=0.145~\rm{GeV}$, which is supported by lattice QCD studies~\cite{Anand:2017kar,HotQCD:2014kol}. We acknowledge that a more rigorous approach would involve dynamically determining the transition temperature and related quantities (such as pressure and latent heat) by modeling the confined (hadronic) phase and matching it with the deconfined phase. For instance, the hadronic phase can be described using the Hadron Resonance Gas  (HRG) approach~\cite{Karsch:2003vd}, possibly restricted to light mesonic components for tractability~\cite{Lerambert-Potin:2021ohy}. Such an analysis lies beyond the scope of the present work and will be considered in future extensions.}. Now, the total peak frequency of SGWB reads
\begin{align}
\label{eq24}
{\nu_{{\rm{total }}}} & ={\nu_{{\rm{BC}}}}+ {\nu_{{\rm{SW}}}}+{\nu_{{\rm{MHD}}}}
\nonumber \\
&= \left( {\frac{{0.62 \alpha }}{{1.8 - 0.1\mu  + {\mu ^2}}} + \frac{{38 \alpha }}{{31\mu }} + \frac{{7 \alpha }}{{4\mu }}} \right)\left( {\frac{{{a_*}}}{{{a_0}}}} \right)
\nonumber \\
&  = \left( {\frac{{0.62 \alpha}}{{1.8 - 0.1\mu  + {\mu ^2}}} + \frac{{38 \alpha}}{{31\mu }} + \frac{{7 \alpha}}{{4\mu }}} \right)\frac{{{T_0}}}{{{T_*}}}{\left[ {\frac{{{g_s}\left( {{T_0}} \right)}}{{{g_s}\left( {{T_*}} \right)}}} \right]^{\frac{1}{3}}}{\left( {\frac{{1 + 16{\beta_0} T_0^2}}{{1 + 16{\beta_0} T_*^2}}} \right)^{\frac{2}{3}}}{\left( {\frac{{1 + 40{\beta_0} T_0^2}}{{1 + 40{\beta_0} T_*^2}}} \right)^{\frac{1}{3}}}.
\end{align}

\begin{figure}[htbp]
\centering
\includegraphics[width=0.59\linewidth]{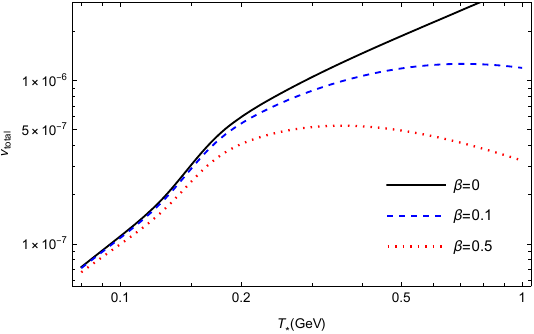}
\caption{The total peak frequency $\nu_{\rm{total}}$ of SGWB at the epoch of phase transition with different values of deformation parameter ${\beta_0}$.}
\label{fig7}
\end{figure}
As shown in figure~\ref{fig7}, the total peak frequency of SGWB exhibits a clear dependence on both the transition temperature $T_*$ and the deformation parameter ${\beta_0}$. In the absence of GUP corrections (${\beta_0} = 0$), $\nu_{\rm{total}}$ remains at its standard value determined by eq. (37) and increases slowly with $T_*$.  As ${\beta_0}$ increases, the peak frequency exhibits a continuous redshift, and the magnitude of this redshift becomes more pronounced for larger ${\beta_0}$. This behavior arises from the enhanced GUP-induced modification to the ${{{a_*}} \mathord{\left/ {\vphantom {{{a_*}} {{a_0}}}} \right. \kern-\nulldelimiterspace} {{a_0}}}$, which reduces the frequency of today relative to its generation value. Consequently, larger values of ${\beta_0}$ shift the SGWB peak to lower frequencies, which could influence future pulsar timing array detections, particularly with high-sensitivity measurements.

\begin{figure*}
\centering
\includegraphics[width=0.75\linewidth]{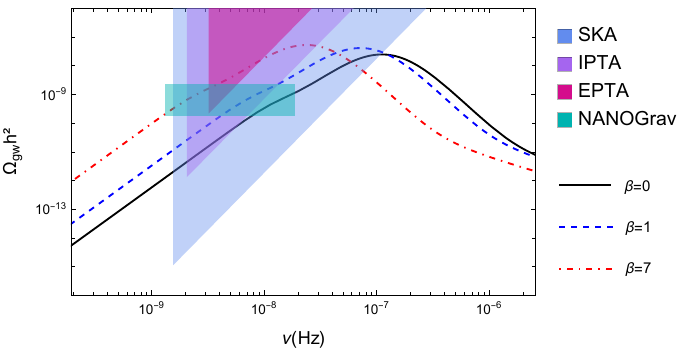}
\caption{The SGWB energy spectra from three sources in QCD phase transition as a function of frequency with different values of deformation parameter ${\beta_0}$.}
\label{fig6}
\end{figure*}

Finally, based on eq.~(\ref{eq18})-eq.~(\ref{eq23}), the total energy density of the SGWB becomes
\begin{align}
\label{eq25}
{\Omega _{{\rm{gw}}}}{h^2} & = \left[ {\Omega _{{\rm{gw}}}^{{\rm{BC}}}\left( \nu \right) + \Omega _{{\rm{gw}}}^{{\rm{SW}}}\left( \nu\right) + \Omega _{{\rm{gw}}}^{{\rm{MHD}}}\left( \nu \right)} \right]{h^2}
\nonumber\\
& = \left[ {\Omega _{{\rm{g}}{{\rm{w}}_*}}^{{\rm{BC}}}\left( \nu \right) + \Omega _{{\rm{g}}{{\rm{w}}_*}}^{{\rm{SW}}}\left( \nu \right) + \Omega _{{\rm{g}}{{\rm{w}}_*}}^{{\rm{MHD}}}\left( \nu \right)} \right]{\Omega _{r0}}{h^2} {\left( {\frac{{{T_r}}}{{{T_*}}}} \right)^4}{\left[ {\frac{{{g_s}\left( {{T_r}} \right)}}{{{g_s}\left( {{T_*}} \right)}}} \right]^{\frac{4}{3}}}\frac{{{g_s}{{\left( {{T_*}} \right)}^{1 + \omega \left( {{T_*}} \right)}}}}{{{g_s}{{\left( {{T_r}} \right)}^{1 + \omega \left( {{T_r}} \right)}}}}
\nonumber\\
& \times {\left( {\frac{{1 + 16{\beta_0} T_r^2}}{{1 + 16{\beta_0} T_*^2}}} \right)^{\frac{8}{3}}}{\left( {\frac{{1 + 40{\beta_0} T_r^2}}{{1 + 40{\beta_0} T_*^2}}} \right)^{\frac{4}{3}}}\exp \left\{ {\int_{{T_r}}^{{T_*}} {\frac{3}{T}} \left( {1 + {\omega _{{\rm{eff}}}}} \right)\left[ {1 + \frac{{16{\beta_0} {T^2}\left( {3 + 80{\beta_0} {T^2}} \right)}}{{1 + 56{\beta_0} {T^2} + 640{{\beta_0} ^2}{T^4}}}} \right]{\rm{d}}T} \right\},
\end{align}
where  $h = H_{0}/100~\rm{km^{-1}s Mpc}$. By substituting eq.~(\ref{eq19})-eq.~(\ref{eq23}) into eq.~(\ref{eq25}), we displays the SGWB spectrum generated by QCD phase transitions with varying values of the parameter ${\beta_0}$ in figure~\ref{fig6}.

In figure~\ref{fig6}, we examine the impact of different values of the deformation parameter ${\beta_0}$ on the SGWB spectrum. First, we need to discuss the range of ${\beta_0}$. As shown in previous studies~\cite{Luo:2023rhk,Luo:2024vdd}, the parameter ${\beta_0}$ of the higher-order GUP~(\ref{eq1}) has been constrained using various theoretical approaches and observational data in the range of $10^{9}$ to $10^{90}$, which leads to the range of deformation parameter ${\beta_0}$ between $10^{-61}$ and $10^{20}$, depending on the specific approach and dataset~\cite{Tawfik:2015rva,Guo:2015ldd}. This wide range offers significant flexibility in exploring phenomenological implications. In this analysis, we therefore consider three representative values. For ${\beta_0}=0$, the black solid curve represents the scenario without GUP corrections, showing a peak around $100$ nHz, detectable by SKA, IPTA, and NANOGrav. However, detectability by IPTA and NANOGrav is marginal, as only a small portion of the spectrum overlaps their sensitivity bands. For the case ${\beta_0}=1$, a representative value commonly adopted in the GUP literature, the blue dashed curve indicates that the peak frequency is significantly shifted downward to around $50$ nHz, substantially improving the detectability by SKA, IPTA, and NANOGrav. This choice provides a clear illustration of how even moderate quantum gravity corrections can enhance the observational prospects. In an intentionally idealized scenario with a relatively large GUP correction (${\beta_0} =7$), the red dot-dashed curve further shifts the peak downward to approximately $10$ nHz, thus becoming optimally detectable by all four considered experiments. This scenario indicates that a larger GUP parameter allows the SGWB signal to be confirmed by a wider range of detectors. Nevertheless, considering theoretical expectations that the GUP parameter should be very small, the most plausible scenario is that SGWB with realistic GUP effects will be detected by SKA, IPTA, and NANOGrav.

\section{Conclusion}
\label{SECTIV}
In this work, we have investigated the SGWB generated from a first-order cosmological QCD phase transition within the framework of a new GUP~(\ref{eq1}). We derived the higher-order GUP corrections to the SGWB spectrum, emphasizing the distinctions and physical implications arising from positive and negative deformation parameters. Our analysis, based on the assumption of a first-order QCD phase transition, demonstrates that only positive GUP parameters yield physically consistent and observationally viable gravitational wave signals, characterized by a notable shift in the SGWB peak toward lower frequencies and a modest enhancement in energy density as the GUP parameter increases. These modifications play a crucial role in understanding the quantum gravity effects during the cosmological evolution. Furthermore, we evaluated the detectability of these modified SGWB signals using current and forthcoming pulsar timing array experiments, such as SKA, IPTA, EPTA, and NANOGrav. Our results reveal that these facilities possess sufficient sensitivity to potentially detect QG corrected gravitational wave signals. Hence, our study not only provides new insights into the interaction between the effects of QG and cosmological phase transitions, but also highlights the promising potential of gravitational wave astronomy as a powerful tool for exploring and testing Planck-scale QG theories.

Despite these findings, some simplifications were made in modeling the QCD phase transition. In particular, a typical critical temperature was used based on previous studies and several phenomenological parameters (such as bubble wall velocity, latent heat, and efficiency factors) were taken as fixed constants. A more self-consistent treatment would involve incorporating realistic QCD modeling, e.g., using the HRG approach, and deriving these parameters dynamically within the GUP-modified thermodynamic framework. Investigating the dependence of these quantities on the GUP parameter could reveal additional QG imprints in the SGWB spectrum. These directions will be explored in future work.

\bibliography{sn-article}
\end{document}